\documentclass[%
 reprint,
nofootinbib,
 amsmath,amssymb,
 aps,
]{revtex4-2}

\usepackage{graphicx}% Include figure files
\usepackage{dcolumn}% Align table columns on decimal point
\usepackage{bm}% bold math
\usepackage{ytableau}

\begin{document}

%\preprint{APS/123-QED}

\title{Spacetime with prescribed hidden symmetry}% Force line breaks with \\

%\thanks{A footnote to the article title}

\author{Song He}
 \email{hesong@jlu.edu.cn}
\affiliation{
 Center for Theoretical Physics and College of Physics, Jilin University,\\
 Changchun 130012, China,
}
\affiliation{School of Physical Science and Technology, Ningbo University,\\ Ningbo 315211, China
}
\affiliation{Max Planck Institute for Gravitational Physics (Albert Einstein Institute),\\ 
Am M\"uhlenberg 1, 14476 Golm, Germany
}
\author{Yi Li}%
 \email{liyi@fudan.edu.cn}
\affiliation{
 Center for Theoretical Physics and College of Physics, Jilin University,\\
 Changchun 130012, China
}

\date{\today}% It is always \today, today,

\begin{abstract}
In this paper, we investigate spacetime characterized by a hidden symmetry defined by a given Killing tensor. To exhibit this hidden symmetry, the inverse metric must commute with the Killing tensor under the Schouten-Nijenhuis bracket, which translates into a system of partial differential equations (PDEs) for the inverse metric. For some significant examples, we solve these PDEs directly, deriving spacetimes with prescribed hidden symmetries, including those specified by higher-rank Killing tensors. Utilizing the hidden symmetries, we study related problems such as null geodesics, photon region, and separation of variables of wave equations. Through this work, we aim to demonstrate that hidden symmetry is more accessible than previously believed.
\end{abstract}

%Use showkeys class option if keyword
\maketitle

%\tableofcontents

\section{Introduction}\label{sec1}

Symmetry is integral to our understanding of nature and serves as a primary tool for obtaining exact results in physical problems. In gravitational theory, the fundamental symmetry of spacetime is isometry, characterized infinitesimally by Killing vectors. First studied in \cite{staeckel1895integration}, a Killing tensor $K$ of rank $r$ generalizes the concept of a Killing vector. It is defined as a symmetric tensor that satisfies the Killing equation.
\begin{align} \label{Killing tensor def eqn}
    \nabla^{(\beta} K^{\alpha_1\ldots\alpha_r)} = 0
\end{align}
It corresponds to a constant of geodesic motion that is a monomial of the momentum and generates symmetry in the phase space that cannot be projected to the position space. We call it ``hidden symmetry'', as opposed to ``explicit symmetry'' characterized by a Killing vector. A symmetrized tensor product of Killing tensors is also a Killing tensor; it's called reducible, and it doesn't generate an independent symmetry in the phase space. We are only interested in irreducible Killing tensors.

The primary example of a Killing tensor is Carter's constant \cite{Carter:1968rr, Carter:1968ks} for geodesic motion in the Kerr black hole, which corresponds to a rank-two Killing tensor \cite{Walker:1970un}. The Kerr spacetime has been generalized to include a cosmological constant, NUT parameters, and electromagnetic charge \cite{Carter:1968ks, carter1968new}, and extended to higher dimensions \cite{myers1986black, Gibbons:2004uw, Gibbons:2004js, Chen:2006xh}. The Kerr family of metrics admits a series of Killing tensors generated from a non-degenerate closed conformal Killing-Yano 2-form \cite{demianski1980separability, carter1987separability, Frolov:2003en, Frolov:2006dqt, Kubiznak:2006kt, Krtous:2006qy, Frolov:2008jr}; see \cite{Yasui:2011pr, Cariglia:2014ysa, Frolov:2017kze} for informative reviews. Beyond the Kerr family, Killing tensors have been identified in Taub-NUT spacetimes \cite{Gibbons:1986hz, Chong:2004hw}, pp waves \cite{Keane:2010hg, Kruglikov:2022udn}, and various supergravity solutions \cite{Chow:2008fe}. In particular, higher-rank ($r \geq 3$) Killing tensors were found in \cite{Gibbons:2011hg, Gibbons:2011nt, Galajinsky:2012vn, Cariglia:2014dfa, Cariglia:2015fva, Dunajski:2024kqo}. These spacetimes were derived via the Eisenhart lift \cite{eisenhart1928dynamical, Duval:1984cj, Duval:1990hj} from integrable systems with constants of motion of higher powers in momentum. More recently, Killing tensors have also been discovered in modified Lense-Thirring spacetimes \cite{Baines:2021qaw, Gray:2021toe} and their higher-dimensional generalizations \cite{Gray:2021toe, Sadeghian:2022ihp}.

For these spacetimes, hidden symmetries are as effective as explicit symmetries in solving physical problems. The integrability of geodesic motion is crucial for the analytic study of bounded orbits \cite{Wilkins:1972rs, Teo:2003ltt, Hackmann:2011wp}, and consequently for the analysis of photon regions, black hole shadows, and photon rings \cite{bardeen1973black, Grenzebach:2014fha, Grenzebach:2015oea, Konoplya:2021slg, Gralla:2019xty}. These are key observational targets for the Event Horizon Telescope \cite{EventHorizonTelescope:2022wkp, EventHorizonTelescope:2022xqj} to test Einstein's theory of gravity. Additionally, hidden symmetries enable the separation of variables in various wave equations \cite{Carter:1968rr, Carter:1968ks, Teukolsky:1972my, Teukolsky:1973ha, Unruh:1973bda, Page:1976jj, Chandrasekhar:1976ap, Frolov:2006pe, Oota:2007vx, Wu:2008df, Oota:2008uj, Kolar:2015hea, Lunin:2017drx, Frolov:2018pys, Krtous:2018bvk, Frolov:2018ezx, Houri:2019nun}. This is valuable for analyzing stability, superradiance, and dark photon detection \cite{andersson2015hidden, Giorgi:2021skz, Dolan:2018dqv, Cayuso:2019ieu, Caputo:2021efm}, as well as in theoretical studies of holography \cite{Guica:2008mu, Castro:2010fd, Compere:2012jk}.

Despite their ubiquitous usefulness, hidden symmetries are often elusive in the literature, with examples being limited and scattered. Numerous tests for hidden symmetries in on-shell geometries have yielded negative results \cite{Owen:2021eez}. Notably, no explicit example of irreducible higher-rank Killing tensors was known until 2011 \cite{Gibbons:2011hg}, and these examples, along with subsequent ones, were reverse-engineered from integrable systems.

In this paper, we aim to demonstrate that hidden symmetries, in some sense, are as accessible as explicit symmetries. Just as we can write down metrics with explicit time translation or spherical symmetry, we can also solve for metrics with prescribed hidden symmetry defined by a Killing tensor. Such a metric, in its upper indices (inverse metric), is a non-degenerate rank-two symmetric tensor that commutes with the given Killing tensor under the Schouten-Nijenhuis bracket. The vanishing of the bracket translates into a system of PDEs for the inverse metric. We begin by reviewing the basics of Killing tensors and hidden symmetries. Subsequently, we compute explicit examples of spacetimes with prescribed hidden symmetries. Then, we apply the hidden symmetries to the analysis of spherical photon orbits, photon region, and separation of variables of wave equations. We conclude with a discussion of open questions and interesting directions for future research.

\section{Killing tensor and hidden symmetry}\label{sec2}

The best way to understand the basic definitions of Killing tensor is by using Hamiltonian mechanics in spacetime. A rank $r$ symmetric tensor $K$ corresponds to a monomial in momentum $K^{\alpha_1\ldots\alpha_r}p_{\alpha_1}\ldots p_{\alpha_r}$, which we abbreviate as $K\circ p$, as a phase space quantity. The Poisson bracket of two monomials is another monomial; hence, it induces the Schouten-Nijenhuis bracket on symmetric tensors
\begin{align}
    [K_1,K_2]_{\text{SN}}\circ p = -\{K_1\circ p,K_2\circ p\}_{\text{Poisson}}
\end{align}
or in components
\begin{align}
    [K_1,K_2]_{\text{SN}}^{\alpha_1\ldots\alpha_{r-1}\gamma\beta_1\ldots\beta_{s-1}} = &r K_1^{\mu(\alpha_1\ldots\alpha_{r-1}}\nabla_\mu K_2^{\gamma\beta_1\ldots\beta_{s-1})} \nonumber\\
    &- s K_2^{\mu(\beta_1\ldots\beta_{s-1}}\nabla_\mu K_1^{\gamma\alpha_1\ldots\alpha_{r-1})}
\end{align}
where $r,s$ are the ranks of $K_1,K_2$ respectively. The Schouten-Nijenhuis bracket is a natural higher-rank generalization of the Lie bracket of vectors, and it doesn't depend on the connection (covariant derivative) chosen despite its appearance in the expression. In particular, we get ordinary derivatives of tensor components if the connection is chosen to be trivial in the coordinate basis. The free Hamiltonian $H=\frac{1}{2}g^{\alpha\beta}p_\alpha p_\beta$ which generates geodesic motion corresponds to the inverse metric $g^\sharp$\footnote{It's customary to use the musical symbol $\sharp$ and $\flat$ to denote type change of tensors by the metric, namely raising and lowering indices.}, and conserved quantities correspond to Killing tensors, namely tensors that satisfy the Killing equation
\begin{align} \label{Killing tensor def by SN}
    [g^\sharp,K]_{\text{SN}} = 0
\end{align}
If we choose a metric-compatible connection, in particular the most often chosen Levi-Civita connection, then (\ref{Killing tensor def by SN}) reduces to (\ref{Killing tensor def eqn}). These two defining equations are equivalent, but (\ref{Killing tensor def eqn}) may belie the simple dependence of the Killing equation on the metric, while (\ref{Killing tensor def by SN}) clearly shows that the Killing equation is linear in the inverse metric. A Killing tensor generates the symmetry in the phase space
\begin{align}
    X_K = \frac{\partial (K\circ p)}{\partial p_\alpha} \partial_{x^\alpha} - \frac{\partial (K\circ p)}{\partial x^\alpha} \partial_{p_\alpha}
\end{align}
Such a symmetry cannot be projected to the position space if the rank of $K$ is greater than one.\\

In a spacetime with a given metric, (\ref{Killing tensor def eqn}) or (\ref{Killing tensor def by SN}) is an overdetermined PDE system (see \cite{dunajski2008overdetermined} e.g.) for Killing tensors with a given rank. All Killing tensors of that rank can be found by the following algorithm \cite{wolf1998structural, Houri:2017tlk}, conveniently using Young symmetrizers (see \cite{Fulton:2004uyc} e.g.), which put tensor indices in a Young tableau and symmetrize within each row then anti-symmetrize within each column. We repeatedly differentiate the defining equation (\ref{Killing tensor def eqn}) and take higher derivatives of $K$ as independent variables (prolongation). Then, we decompose them into different symmetries corresponding to different Young tableaux. We find the independent prolongation variables are
\begin{align}
    Y_{\tiny \begin{ytableau}
        \alpha_1 & \ldots & \ldots & \alpha_r \\
        \beta_1 & \ldots & \beta_q
    \end{ytableau}} \nabla^{\beta_q}\ldots\nabla^{\beta_1} K^{\alpha_1\ldots\alpha_r}
\end{align}
with $0\leq q\leq r$. Other terms either vanish by the defining equation (\ref{Killing tensor def eqn}), for example,
\begin{align}
    Y_{\tiny \begin{ytableau}
        \alpha_1 & \ldots & \ldots & \alpha_r & \beta_1 \\
        \beta_2 & \ldots & \beta_q
    \end{ytableau}} \nabla^{\beta_q}\ldots\nabla^{\beta_1} K^{\alpha_1\ldots\alpha_r}
\end{align}
or reduce to lower order derivatives of $K$ by anti-symmetrizing indices of derivatives, for example
\begin{align}
    Y_{\tiny \begin{ytableau}
        \alpha_1 & \ldots & \ldots & \alpha_r \\
        \beta_1 & \ldots & \beta_q \\
        \beta_2
    \end{ytableau}} \nabla^{\beta_q}\ldots\nabla^{\beta_1} K^{\alpha_1\ldots\alpha_r}
\end{align}
The procedure of prolongation ends at $q=r$ when further differentiation yields no higher derivatives\footnote{In the language of Young tableau, the second row is full.}, but a linear relation between the independent prolongation variables obtained. We then repeatedly differentiate this linear relation to get new linear relations. Because of the finite dimension, after finitely many steps, differentiation no longer generates new linear relation that is linearly independent, then Killing tensors are computed by solving all the linear relations we have obtained to this point.

The theme of this paper is to find a metric with prescribed hidden symmetry by solving for $g^\sharp$ from (\ref{Killing tensor def by SN}) with a given $K$. If $K$ is of rank two and non-degenerate (or can be made non-degenerate by adding other Killing tensors in the spacetime), an algorithm similar to the one discussed above works by swapping the roles of $g^\sharp$ and $K$\footnote{The symmetry between the inverse metric and a non-degenerate rank two Killing tensor has been noted in \cite{Rietdijk:1995ye}.}. If $K$ is of rank two but degenerate, we can choose a basis in which $K$ takes the canonical form of a symmetric matrix, specifically a diagonal matrix with eigenvalues $\pm1,0$. In this basis, we choose the connection to be trivial, making it $K$-compatible, namely the covariant derivative of $K$ is zero. With this connection, the Killing equation (\ref{Killing tensor def by SN}) transforms to
\begin{align}
    K^{\mu(\alpha}\nabla_\mu g^{\beta\gamma)} = 0.
\end{align}
Now a similar algorithm works by repeatedly differentiating with $K^{\alpha\mu}\nabla_\mu$. The only thing to check is that the commutator $[K^{\alpha\mu}\nabla_\mu, K^{\beta\nu}\nabla_\nu]$ reduces to a linear action by the Riemann tensor, due to the $K$-compatibility of the connection. For generic Killing tensors, it remains unknown to us how to solve the system of PDEs. In the examples provided below, the prescribed Killing tensor is simple enough that (\ref{Killing tensor def by SN}) can be solved directly.

\section{Examples of spacetime with prescribed hidden symmetry}\label{sec3}

The typical physical scenario for a spacetime with prescribed hidden symmetry is as follows. We begin with a spacetime possessing strong explicit symmetry and then deform the metric to break some of these explicit symmetries, such as rotating a spherical black hole. Our focus is on deformed metrics that preserve a (reducible) Killing tensor of the original metric while breaking the explicit symmetries constituting the Killing tensor. The Killing tensor becomes irreducible and characterizes the hidden symmetry of the deformed metric. Despite the reduced explicit symmetry, the hidden symmetry allows for the exact solution of many problems in the deformed spacetime.

\subsection{Hidden symmetry with broken spherical symmetry}\label{subsec3.1}
Consider a static spherically symmetric spacetime with the metric, which includes all kinds of spherical black holes
\begin{align}
    g^{(0)} = - f(r)dt^2 + \frac{1}{h(r)} dr^2 + r^2 (d\theta^2 + \sin^2\theta d\phi^2)
\end{align}
We have the Killing vectors $L_x=\sin\phi\partial_\theta + \cot\theta\cos\phi\partial_\phi,\;L_y=\cosh\phi\partial_\theta-\cot\theta\sin\phi\partial_\phi,L_z=\partial_\phi$ corresponding the spherical symmetry and $\partial_t$ corresponding to the time translation symmetry. We have the ``angular momentum squared'' as a reducible rank two Killing tensor
\begin{align}
    K = L_x^2 + L_y^2 + L_z^2 = \partial_\theta^2 + \frac{1}{\sin^2\theta}\partial_\phi^2
\end{align}
Now we consider a deformed metric for which the spherical symmetry is broken, but the Killing tensor $K$, $\partial_t$, and $\partial_\phi$ are preserved. Then, including the inverse metric itself, we have four commuting Killing tensors, preserving the integrability of geodesic motion. Looking for a solution to (\ref{Killing tensor def by SN}) that is independent on $t,\phi$, we find
\begin{align}
    (g^{\mu\nu}) = \begin{pmatrix}
        G^{tt}(r) & G^{tr}(r) & 0 & G^{t\phi}(r)\\
        G^{tr}(r) & G^{rr}(r) & 0 & G^{r\phi}(r)\\
        0 & 0 & G^{\theta\theta}(r) & 0 \\
        G^{t\phi}(r) & G^{r\phi}(r) & 0 & \frac{G^{\theta\theta}(r)}{\sin^2\theta} + H^{\phi\phi}(r)
    \end{pmatrix}
\end{align}
Meanwhile, an infinitesimal diffeomorphism that preserves $\partial_t$, $\partial_\phi$ and $K$ is of the form
\begin{align}
    V = V^t(r)\partial_t + V^r(r) \partial_r + V^\phi(r) \partial_\phi
\end{align}
Modulo variation of the inverse metric induced by such a diffeomorphism, the general deformed inverse metric that preserves $\partial_t$, $\partial_\phi$ and $K$ but breaks the spherical symmetry, is given by
\begin{align}
    (g^{\mu\nu}) = \begin{pmatrix}
        -\frac{1}{f(r)} & 0 & 0 & \frac{\omega(r)}{f(r)}\\
        0 & h(r) & 0 & 0\\
        0 & 0 & \frac{1}{r^2} & 0 \nonumber\\
        \frac{\omega(r)}{f(r)} & 0 & 0 & \frac{r^2}{\sin^2\theta} + \frac{s(r)}{r^2}+\frac{\omega(r)^2}{f(r)}
    \end{pmatrix}
\end{align}
The form we express the inverse metric in terms of the parametric functions $\omega(r),s(r)$ was chosen for a simple expression of the metric itself
\begin{align}
    g = -f(r)dt^2+\frac{1}{h(r)}dr^2 + r^2d\theta^2 + \frac{r^2\sin^2\theta}{1+s(r)\sin^2\theta} (d\phi+\omega(r)dt)^2
\end{align}

Our result generalizes the modified Lense-Thirring spacetime in \cite{Baines:2021qaw,Gray:2021toe}\footnote{The metric in \cite{Baines:2021qaw} has the additional cross term $dtdr$, which in our formalism can be gauged away by diffeomorphism.}. We do not restrict the form of $f(r),g(r)$ to start with; in addition, the ``angular velocity'' $\omega$ is allowed to be an arbitrary function of $r$, and most importantly, a ``squash'' characterized by $s(r)$ is permitted.

As discussed in the introduction, hidden symmetry finds many applications. Here we consider spherical null geodesics\footnote{See also \cite{Baines:2021qfm} for the study of geodesics in the particular modified Lense-Thirring spacetime.} and the photon region following the spirit of \cite{Teo:2003ltt, Grenzebach:2014fha}. We have four constants of motion corresponding to $\partial_t,\partial_\phi,K$ and mass squared which is zero for photon
\begin{align} \label{constants of motion}
    &-f(r)\dot{t}+\omega(r) + \frac{\omega(r)r^2(\omega(r)\dot{t}+\dot{\phi})}{\csc^2\theta+s(r)} = -E \nonumber\\
    &\frac{r^2(\omega(r)\dot{t}+\dot{\phi})}{\csc^2\theta+s(r)} = J \nonumber\\
    &r^4\dot{\theta}^2 + \csc^2\theta \frac{r^4(\omega(r)\dot{t}+\dot{\phi})^2}{(\csc^2\theta+s(r))^2} = C \nonumber\\
    &-f(r)\dot{t}^2 + \frac{\dot{r}^2}{h(r)} + \frac{r^2\big( (\csc^2\theta+s(r))\dot{\theta}^2 + (\dot{\phi}+\omega(r)\dot{t})^2\big)}{\csc^2\theta+s(r)} = 0
\end{align}
We use the first three equations to eliminate $\dot{t},\dot{\phi},\dot{\theta}$ in favor of the constants energy $E$, $z$-axis angular momentum $J$ and ``angular momentum squared'' $C$, and plug in the last equation, we find
\begin{align}
   \frac{\dot{r}^2}{E^2} = h(r)\big( \frac{(1+\mathcal{J}\omega(r))^2}{f(r)} - \frac{\mathcal{C}+\mathcal{J}^2 s(r)}{r^2} \big)
\end{align}
where we have defined $\mathcal{J}=\frac{J}{E}$ and $\mathcal{C}=\frac{C}{E^2}$. For spherical orbits, we have $\dot{r}=0$ and $\Ddot{r}=0$. So we have the equations determining the radius of the spherical orbits in terms of $\mathcal{J},\mathcal{C}$
\begin{align} \label{orbit radius constants of motion}
    h(r)\big( \frac{(1+\mathcal{J}\omega(r))^2}{f(r)} - \frac{\mathcal{C}+\mathcal{J}^2 s(r)}{r^2} \big) &= 0 \nonumber\\
    \frac{d}{dr} \big[ h(r)\big( \frac{(1+\mathcal{J}\omega(r))^2}{f(r)} - \frac{\mathcal{C}+\mathcal{J}^2 s(r)}{r^2} \big) \big] &= 0
\end{align}
In turn, we express $\mathcal{J},\mathcal{C}$ in terms of the radius and plug in the inequality from the third equation in (\ref{constants of motion})
\begin{align} \label{photon region}
    \mathcal{C}-\csc^2\theta \mathcal{J}^2 = \frac{r^4\dot{\theta}^2}{E^2} \geq 0
\end{align}
This inequality gives us the region where spherical photon orbits are allowed, namely the photon region. Taking the modified Lense-Thirring spacetime \cite{Gray:2021toe} for an example, we have $f(r)=h(r)=1-\frac{2M}{r}+\frac{q^2}{r^2}+\frac{r^2}{l^2},\omega(r)=a \frac{f(r)-1}{r^2}$ and $s(r)=0$. By (\ref{orbit radius constants of motion}) $\mathcal{J},\mathcal{C}$ are given by the spherical orbit radius $r$ as
\begin{widetext}
\begin{align}
    \mathcal{J} &= -\frac{l^2 r^4\big(r(r-3M)+2q^2\big)}{a\big[ r^4\big(r(r+3M)-2q^2\big) + l^2\big(2Mr^2(2r-3M)-q^2r(3r-7M)-2q^4\big) \big]} \nonumber\\
    \mathcal{C} &= \frac{4l^2r^4(3Mr-2q^2)^2\big(r^4+l^2(q^2+r(r-2M))\big)}{\big[ r^4(r(r+3M)-2q^2) + l^2\big(2Mr^2(2r-3M)-q^2r(3r-7M)-2q^4\big) \big]^2}
\end{align}
\end{widetext}
The photon region is given by (\ref{photon region}). It takes a simple form for the asymptotically flat and uncharged case ($l\to\infty,q=0$) 
\begin{align}
    (\frac{r}{M})^3(\frac{r}{M}-3)^2 \leq \frac{36a^2}{M^2}(\frac{r}{M}-2) \sin^2\theta
\end{align}

\subsection{Hidden symmetry with broken planar symmetry}\label{subsec3.2}

Consider a static plane-symmetric spacetime with the metric
\begin{align}
    g^{(0)} &= -f(r)dt^2 + \frac{1}{h(r)}dr^2 + r^2 (d\rho^2 + \rho^2 d\phi^2) \nonumber\\
    &= -f(r)dt^2 + \frac{1}{h(r)}dr^2 + r^2 (dx^2+dy^2)
\end{align}
We have Killing vectors $\partial_x,\partial_y,\partial_\phi$ corresponding to the planar symmetry and $\partial_t$ corresponding to the time translation symmetry. We have the ``spatial momentum squared'' as a reducible rank two Killing tensor
\begin{align}
    K=\partial_x^2+\partial_y^2=\partial_\rho^2 + \frac{1}{\rho^2}\partial_\phi^2
\end{align}
Similar to the spherical case, we can show that the general deformed metric (modulo diffeomorphism) that preserves the Killing tensor $K$, $\partial_t$ and $\partial_\phi$ but breaks the planar symmetry is
\begin{align}
    ds^2 = -f(r)dt^2 + \frac{1}{h(r)}dr^2 + r^2 d\rho^2 + \frac{r^2\rho^2}{1+s(r)\rho^2} (d\phi+\omega(r)dt)^2
\end{align}\\

As an application of the hidden symmetry in the deformed spacetime, we study the separation of variables of the Klein-Gordon equation. Formally, the separation of variables can be characterized by a set of commuting operators, including the one in the PDE, whose common eigenfunctions span the whole solution space to the PDE. For commuting Killing vectors and rank two Killing tensors, the operators constructed as
\begin{align}
    \mathcal{V} = V^\alpha \nabla_\alpha, \mathcal{K} = \nabla_\alpha K^{\alpha\beta} \nabla_\beta
\end{align}
are found to commute except for ``anomaly'' terms \cite{Carter:1977pq,Kolar:2015cha}. In our case, we have four operators constructed from the Killing tensors, $\nabla_\alpha\nabla^\alpha$ in the Klein-Gordon equation, $\mathcal{K} = \nabla_\alpha K^{\alpha\beta} \nabla_\beta$ from the rank two Killing tensor, $\partial_t$ and $\partial_\phi$. They mutually commute except for the possible anomaly in $[\nabla_\alpha\nabla^\alpha,\mathcal{K}]$, which vanishes only when $s(r)=s$ is a constant. In this case, the separation of variables is relatively straightforward. We write a solution in the form
\begin{align} \label{KG separation of variables}
    \Phi(t,r,\rho,\phi) = e^{-i E t + i J \phi} \mathcal{F}(r)\mathcal{G}(\rho)
\end{align}
and the Klein-Gordan equation
\begin{align} \label{KG eqn}
    (\nabla^\mu\nabla_\mu-m^2)\Phi = 0
\end{align}
is reduced to
\begin{align}
    &\frac{r^2 \big (h^{'}(r) + h(r)(\frac{4}{r}+\frac{f^{'}(r)}{f(r)}) \big) \mathcal{F}^{'}(r)}{2\mathcal{F}(r)} + \frac{r^2 h(r)\mathcal{F}^{''}(r)}{\mathcal{F}(r)} \nonumber\\
    &+ \frac{r^2(E+J \omega(r))^2}{f(r)} - m^2 r^2 \nonumber\\
    & + \frac{\mathcal{G}^{''}(\rho)}{\mathcal{G}(\rho)} + \frac{\mathcal{G}^{'}(\rho)}{\mathcal{G}(\rho)\rho(1+s\rho^2)} - J^2 \frac{(1+s\rho^2)}{\rho^2} = 0
\end{align}
Usually, the separation of variables above results in Sturm-Liouville eigenfunction equations for $\mathcal{F},\mathcal{G}$. The eigenfunctions form a complete basis and justify the separation of variables method to solve the original linear PDE, namely solutions of the form (\ref{KG separation of variables}), with variables separated, span the space of solutions to the Klein-Gordon equation. Now, we consider the case when $s(r)$ is an arbitrary function. It's convenient to change the coordinate $v=\frac{1}{\rho^2}$ and the metric now reads
\begin{align}
    g = -f(r)dt^2 + \frac{1}{h(r)} dr^2 + \frac{r^2}{4v^3}dv^2 + \frac{r^2}{v+s(r)} d\phi^2
\end{align}
Again, we write a solution in the form with variables separated
\begin{align} \label{KG separation of variables 1}
    \Phi(t,r,v,\phi) = e^{-i E t + i J \phi} \mathcal{F}(r)\mathcal{G}(v)
\end{align}
The Klein-Gordon equation is reduced to
\begin{align}
    a_1(r)+b_1(v)+(v+s(r))(a_2(r)+b_2(v)) = 0
\end{align}
where we define
\begin{widetext}
\begin{align}
    a_1(r) &= -\frac{r^2h(r)s^{'}(r)\mathcal{F}^{'}(r)}{\mathcal{F}(r)} \nonumber\\
    b_1(v) &= -\frac{4v^3\mathcal{G}^{'}(v)}{\mathcal{G}(v)}\nonumber\\
    a_2(r) &= \frac{2E^2r^2}{f(r)} - 2m^2r^2 -2J^2s(r) + \big(4rh(r)+\frac{r^2h(r)f^{'}(r)}{f(r)} + r^2h^{'}(r)\big) \frac{\mathcal{F}^{'}(r)}{\mathcal{F}(r)} + \frac{2r^2h(r)\mathcal{F}^{''}(r)}{\mathcal{F}(r)} \nonumber\\
    b_2(v) &= -2J^2v + \frac{12v^2\mathcal{G}^{'}(v)}{\mathcal{G}(v)} + \frac{8v^3\mathcal{G}^{''}(v)}{\mathcal{G}(v)}
\end{align}
Differentiating with respect to $v$, we find
\begin{align}
    b_1^{'}(v)+a_2(r)+b_2(v)+v b_2^{'}(v) + s(r)b_2^{'}(v) = 0
\end{align}
We have assumed $s(r)$ is not a constant function, so $b_2^{'}(v)$ must be a constant, say, $C_1$. Then $a_2(r)+C_1 s(r)$ must be another constant, say, $C_2$. A bit more computation leads us to the following separation of variables
\begin{align}
    &b_2(v) = C_1 v + C_3, \quad b_1(v) = -C_1 v^2 - (C_2+C_3) v + C_4 \nonumber\\
    &a_2(r) + C_1 s(r) = C_2, \quad a_1(r) + (C_2+C_3) s(r) - C_1 s(r)^2 + C_4 = 0
\end{align}
\end{widetext}
where $C_3$ and $C_4$ are another two constants. These four ODEs, two second order and two first order, are clearly over-constraining compared to the original PDE. For example, from the second equation above and the boundary condition that $\mathcal{G}(v)$ must be finite as $v\to 0$ and $v\to\infty$, we find $\mathcal{G}(v)$ is a constant. Therefore, solutions of the form with variables separated only span a very small subspace of the whole solution space of the Klein-Gordon equation, so separation of variables in terms of $t,r,\rho,\phi$ coordinates is not possible. We knew $\nabla_\alpha\nabla^\alpha$ doesn't commute with $\mathcal{K}$ if $s(r)$ is not a constant, our direct computation ensures that there is also no modification to $\mathcal{K}$ to make the Klein-Gordon equation separable in $t,r,\rho,\phi$ coordinates.

We have relatively few examples in the literature of irreducible higher rank Killing tensors. In fact, here we can solve for a deformed metric that preserves higher rank ($r\geq 3$) Killing tensors, for example 
\begin{align}
    T= K\partial_t =\partial_t(\partial_x^2+\partial_y^2)
\end{align}
Looking for a solution to (\ref{Killing tensor def by SN}) that is independent on $\phi$, we find
\begin{widetext}
\begin{align}
    (g^{\mu\nu}) = \begin{pmatrix}
        G^{tt}(r)+4t^2a(r)-4tb(r) & G^{tr}(r)-2tG^{r\rho}(r)t & \rho(-2ta(r)+b(r)) & \omega(r)-2tc(r) \nonumber\\
       G^{tr}(r)-2tG^{r\rho}(r)t & G^{rr}(r) & \rho G^{r\rho}(r) & G^{r\phi}(r) \nonumber\\
       \rho(-2ta(r)+b(r)) & \rho G^{r\rho}(r) & G^{\rho\rho}(r)+\rho^2a(r) & \rho c(r) \nonumber\\
      \omega(r)-2tc(r) & G^{r\phi}(r) & \rho c(r)  & \frac{G^{\rho\rho}(r)}{\rho^2}+s(r)
    \end{pmatrix}
\end{align}
An infinitesimal diffeomorphism that preserves $\partial_\phi$ and $T$ is of the form
\begin{align}
    V = (-2t W^\rho(r) + W^t(r))\partial_t + V^r(r)\partial_r + \rho W^\rho(r)\partial_\rho + V^\phi(r) \partial_\phi
\end{align}
Modulo this diffeomorphism, the general deformed inverse metric that preserves $T$ and $\partial_\phi$ but breaks $\partial_t$ or $K$ (thus making $T$ irreducible) is given by
\begin{align}
    (g^{\mu\nu}) = \begin{pmatrix}
        -\frac{1}{f(r)}+4t^2a(r)-4tb(r) & 0 & \rho(-2ta(r)+b(r)) & -2tc(r) \nonumber\\
       0 & h(r) & 0 & 0 \nonumber\\
       \rho(-2ta(r)+b(r)) & 0 & \frac{1}{r^2}+\rho^2a(r) & \rho c(r) \nonumber\\
      -2tc(r) & 0 & \rho c(r)  & \frac{1}{r^2\rho^2}
    \end{pmatrix}
\end{align}
The metric itself takes a complicated form. It becomes simpler if we look at a single mode of deformation, for example, with $a(r),b(r)$ set to zero and only $c(r)$ turned on
\begin{align}
    g = -f(r) dt^2 + \frac{1}{h(r)} dr^2 + r^2 d\rho^2 +\frac{r^2\rho^2}{1+c(r)^2(4t^2r^2\rho^2f(r)-r^4\rho^4)} \big( d\phi - c(r)(2tf(r)dt + r^2\rho d\rho) \big)^2
\end{align}
\end{widetext}

\section{Discussion}\label{sec4}
In this paper, we study spacetimes with prescribed hidden symmetries and provide several examples. Beyond our examples, numerous spacetimes with hidden symmetries can be constructed in a similar manner and used to explore general questions related to hidden symmetry, such as the separation of variables in wave equations or the existence of photon surfaces \cite{Koga:2020akc, Kobialko:2021aqg}. An especially interesting question is how hidden symmetries can be realized for wave equations.

Our construction of spacetime is based solely on symmetry, without reference to any specific physical gravity theory. It is of great interest to investigate whether these spacetimes with hidden symmetries are approximate or exact solutions to relevant gravity theories, with or without matter; see \cite{Gray:2021roq} for work in this direction.

Mathematically, solving the defining equation (\ref{Killing tensor def by SN}) is central to our construction. It is particularly important to develop a qualitative understanding of this system of PDEs when the Killing tensor is of higher rank ($r \geq 3$).

Finally, our work can be extended to other Killing objects related to hidden symmetry, such as the Killing-Yano form \cite{yano1952some, bochner1949curvature} (see also the reviews \cite{Cariglia:2014ysa, Frolov:2017kze}) and the generalized Killing tensor \cite{Aoki:2016ift}.\\

\section*{Acknowledgments}
We would like to thank Tsuyoshi Houri and Hao Ouyang for fruitful discussions. S. H. acknowledges financial support from the Max Planck Partner Group, the Fundamental Research Funds for the Central Universities, and the Natural Science Foundation of China Grants No. 12075101 and No. 12235016.

\nocite{*}

\bibliography{reference}% Produces the bibliography via BibTeX.

\end{document}